\documentclass[twoside]{article}
\usepackage{fleqn,espcrc2}
\usepackage{graphicx}

\newcommand{\AmS}{{\protect\the\textfont2
   A\kern-.1667em\lower.5ex\hbox{M}\kern-.125emS}}
\usepackage{graphicx}
\usepackage{here}
\def\beq{\begin{equation}}
\def\eeq{\end{equation}}
\def\bea{\begin{eqnarray}}
\def\eea{\end{eqnarray}}
\def\bq{\begin{quote}}
\def\eq{\end{quote}}

\def\bear{\begin{array}}
\def\eear{\end{array}}
\def\nnb{\nonumber}
\def\ga{\left(}
\def\dr{\right)}

\def\rar{\rightarrow}
\def\Lrar{\Longrightarrow}

\def\nnb{\nonumber}
\def\la{\langle}
\def\ra{\rangle}
\def\nin{\noindent}
\def\ba{\begin{array}}
\def\ea{\end{array}}

\def\b{\bullet}

\def\mb{\overline{m}}

\def\gam5{\gamma_5}

\title{\bf{QCD Tests of the Puzzling Scalar Mesons
} 
}
\author{ Stephan Narison\address{Laboratoire de Physique Th\'eorique et Astrophysiques,\\
Universit\'e de Montpellier II
Place Eug\`ene Bataillon,
34095 - Montpellier Cedex 05, France.
\\ E-mail:
snarison@yahoo.fr}}
\begin{document}
\pagestyle{plain}
\begin{abstract}
\noindent
Motivated by several recent data, we test the QCD spectral sum rules (QSSR) predictions based on different proposals ($\bar
qq$,
$\bar q\bar q qq$, and gluonium) for the nature of scalar mesons. In the $I=1$ and 1/2 channels, the unusual {\it wrong} splitting
between the
$a_0(980)$ and $\kappa(900)$ and the $a_0(980)$ width can be understood from QSSR within a $\bar qq$ assignement. 
However, none of the $\bar qq$ and $\bar q\bar q qq$ results can explain the large $\kappa$ width, which may suggest that
it can result from a strong interference with non-resonant backgrounds. In the $I=0$ channel, QSSR and some
low-energy theorems (LET) require the existence of a low mass gluonium $\sigma_B$(1 GeV) coupled strongly to Goldstone boson
pairs which plays in the $U(1)_V$ channel, a similar role than the $\eta'$ for the value of the $U(1)_A$ topological charge.
The observed
$\sigma(600)$ and
$f_0(980)$ mesons result from a maximal mixing between the gluonium $\sigma_B$ and $\bar qq$(1 GeV) mesons, a mixing scheme
which passes several experimental tests. 
OZI
violating
$J/\psi\rar
\phi\pi^+\pi^-$,
$D_s\rar 3\pi$ decays and $J/\psi\rar \gamma S$ glueball filter processes may indicate that 
the $f_0(1500),~f_0(1710)$ and
$f_0(1790)$ have significant gluonium component in their wave functions, while  the $f_0(1370)$ is mostly $\bar qq$. Tests of these results
can be provided by the measurements of the pure gluonium $\eta'\eta$ and $4\pi$ specific $U(1)_A$ decay channels.
\vspace*{2mm}
\noindent
\end{abstract}
\maketitle
\section{Introduction}
\nin
The nature of scalar mesons continues to be an intriguing problem in QCD.
Experimentally, there are well established scalar mesons with isospin $I=1$, the 
$a_0(980)$, $a_0(1450)$ with isospin $I=1/2$, the
$K^*_0(1410)$ meson,  and with isospin $I=0$, the $f_0$-mesons at 
980, 1370 \cite{PDG,MONTANET} and 1500 MeV from GAMS, CRYSTAL BARREL \cite{PDG,MONTANET}, WA102 \cite{WA102} and BES \cite{BES}.
Besides these resonances, there are different experimental indications \cite{MONTANET}, especially from BES
\cite{BES}, E791 \cite{BEDIAGA}, FOCUS \cite{MASS}, KLOE \cite{KLOE}, SND \cite{SND}, CMD2 \cite{CMD2}, BELLE
\cite{BELLE}, WA102 \cite{WA102} and $\pi\pi$ scattering data \cite{PIPI,ANISO} for
some other scalar states, with 
$I=0$, the
$\sigma(600)$, $f_0(1710)$ and $f_0(1790)$, and with
$I=1/2$, the
$\kappa(840)$.
The real quark and/or gluon contents of these states are not fully understood, which the
interpretation using effective theories most of them based on a linear realization of chiral symmetry
cannot clarify. In the following, we shall focus on the tests
of the
$\bar qq$, $\bar q\bar q qq$ and gluonium natures of these scalar mesons by confronting the recent
experimental data with some QCD predictions based on QCD spectral sum rules (QSSR)
complemented with some low-energy theorems (LET) \cite{SNB,SNG1,SNG,SNG0} and lattice calculations \cite{PEARDON,KYOTOG}.
\section{The $I=1,~1/2$ scalar mesons}
\nin
{\bf The \boldmath$a_0(980)$ and $\kappa(840)$ masses}\\
These channels are expected to be simpler as we do not expect to have any mixing with
a gluonium. If one assumes that these states are $\bar qq$ mesons, one can naturally
associate them to the divergence of the vector currents:
\bea
a_0(980)&\rar&\partial_\mu V^\mu_{\bar ud}\equiv (m_u-m_d):\bar u(i)d:~,\nnb\\
\kappa(840)&\rar&\partial_\mu V^\mu_{\bar us}\equiv (m_u-m_s):\bar u(i)s:~.
\eea 
Within the QSSR approach, the properties and implications of these mesons can be studied
from the two-point correlator:
\beq
\psi_{\bar uq}(q^2)=i\int d^4x e^{iqx}\la 0|{\cal T}\partial_\mu V^\mu_{\bar
uq}(x)\partial_\mu V^\mu_{\bar uq}(0)^\dagger |0\ra .
\eeq
\begin{figure*}[hbt]
\begin{center}
\includegraphics[width=5cm]{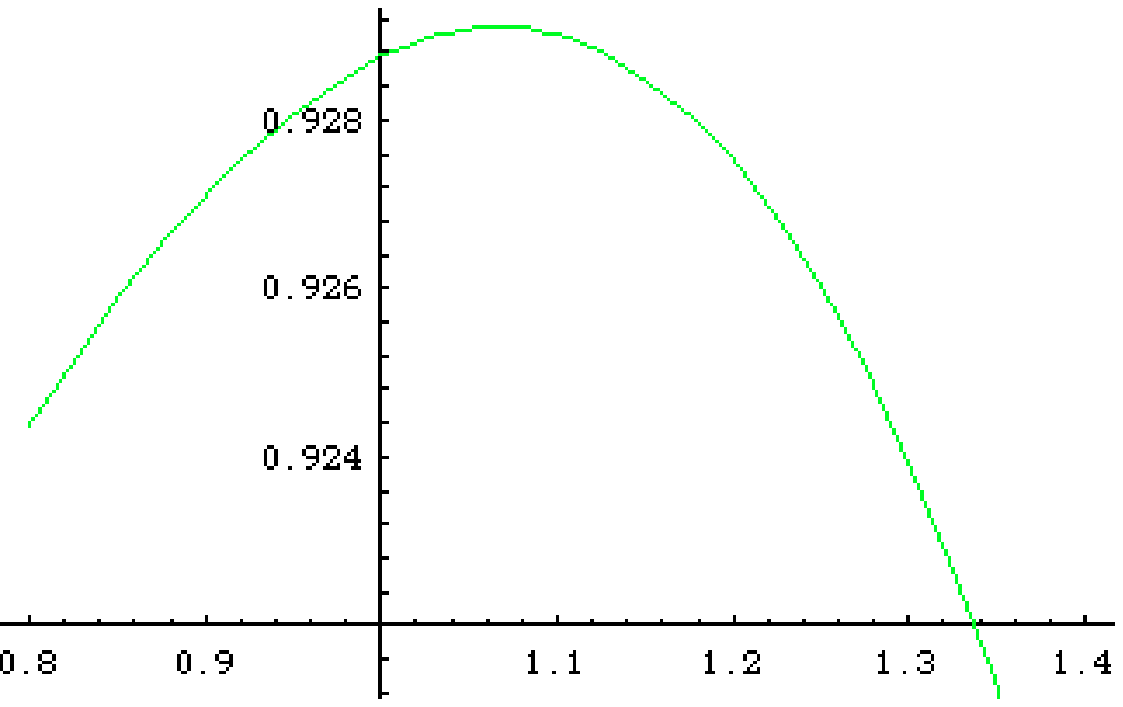}
\includegraphics[width=5cm]{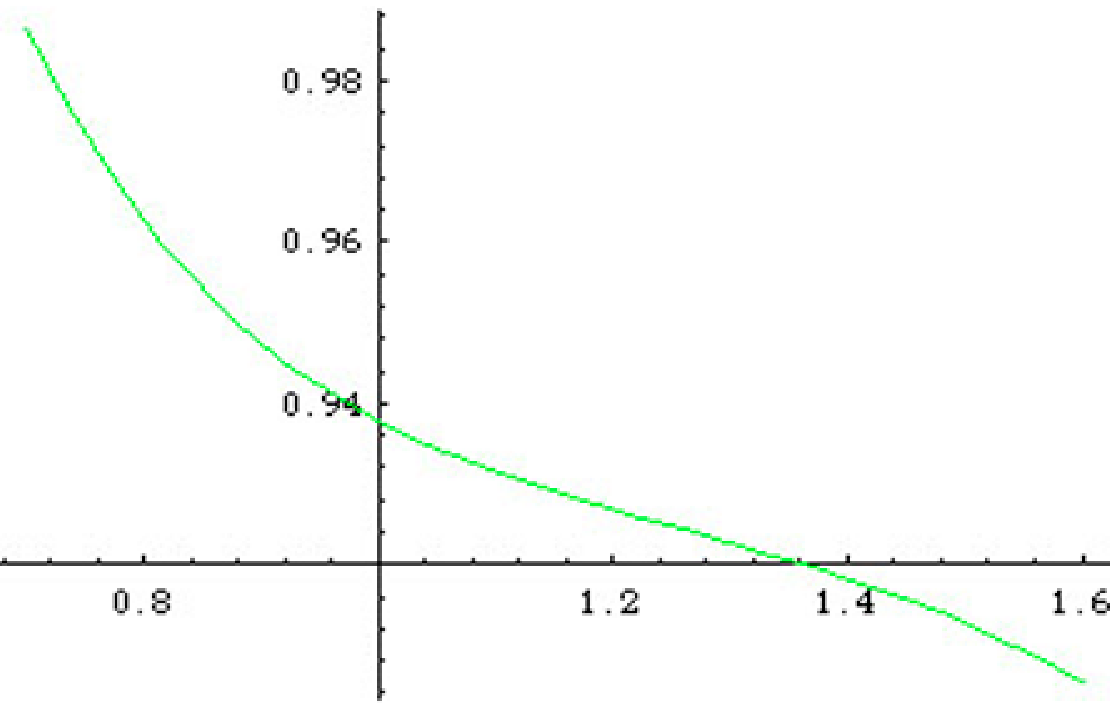}
\caption{$\tau$ in GeV$^{-2}$-dependence of the a) $M_{f(a_0)}$ in GeV and b)
$M_{\kappa}$ in GeV for $\overline m_s$(2 GeV)=98 MeV at a given value of continuum threshold
stability point: $t_c\approx 1$ GeV$^2$.} 
\label{fig: scalar1}
\end{center}
\end{figure*}
\nin
Since the pioneering work in \cite{SCAL}, numerous authors have used and improved
the analysis of the previous correlator for the extraction of the running $u$-$d$ 
mass-difference  and strange quark masses \cite{SNB,SNL}. The improvements come from the
inclusion of higher order terms in the PT QCD series \cite{CHET05,SNB}; the inclusion of the
$1/q^2$ term \cite{CNZ,ZAK} which mimics the UV renormalon effects and which is
also an alternative to the  direct instanton effects where the later is not under a good
quantitative control  due to the incertainties of the instanton
size and widths,...; the treatment of the spectral function using new $K\pi$ phase shift data \cite{OLLER}. 
The obtained value of the $u$-$d$ mass-difference is consistent with some other determinations, 
while the extraction of the
$a_0(980)$ mass from the sum rule \cite{SNB,SNL} is perfectly consistent with the data. Recent analysis
\cite{OLLER} has lead to a value of the strange quark running mass which is consistent with some other
QSSR determinations from different channels \cite{SNMS05}. All these features support the $\bar qq$
meson assignements for the $a_0(980)$ and the $\kappa(840)$ mesons. However, the apparent {\it wrong}
splitting of the $a_0(980)$ and the $\kappa(840)$ mesons is intriguing. Here, we investigate this
analysis using the ratio of exponential Laplace/Borel sum rules:
\beq
{\cal R}(\tau)=-{d\over d\tau}\log{\int_0^\infty dt~e^{-t\tau} {\rm Im}\psi_{\bar ud}(t)}~.
\eeq
Using the PT series to order $\alpha_s^3$, including the NP condensates of dimension 6 and the new
1/$q^2$ terms, we give the prediction for the $a_0(980)$ in Fig. \ref{fig: scalar1}a)  and
for the $\kappa(840)$ in Fig. \ref{fig: scalar1}b) using a NWA. We use the most recent value of $\overline{m}_s$(2
GeV)=$(96\pm 5)$ MeV compiled in \cite{SNMS05}. 
One can see from these figures that the method
reproduces the {\it wrong} splitting of the two mesons. The reason is that the $SU(3)$ breaking effects
increase the value of the sum rule optimization scale compared to the one of the $a_0$ and then emphasize
the contribution of the dimension six condensates, which is a vital correction in the analysis.
Indeed, from the analytic expression of the sum rules, one can qualitatively extract the approximate mass
formula \footnote{Notice that analogous
formula in the vector channel explain with a relatively good accuracy the well-known
$\phi$--$\rho$ and
$K^*$--$\rho$ mass splittings \cite{SNGMO}.}:
\bea
M^2_{\kappa}&\simeq& M^2_{a_0}+2\overline{m}_s^2- 8\pi^2m_s\la\bar ss\ra\tau_0\nnb\\
&+&{3\over 2} {1408\over 81}\pi^3\rho\alpha_s\ga \la \bar ss\ra^2-\la\bar uu\ra^2\dr\tau^2_0\nnb\\
&-&{1\over
3}M^2_\kappa\Gamma_\kappa^2\tau_0,
 \eea
where all different
parameters including the $a_0$ mass  are evaluated at the sum rule optimization scale $\tau_0\simeq 0.8 {\rm GeV}^{-2}$; $\rho\simeq
2$ \cite{LAUNER} indicates the deviation from the vacuum saturation of the four-quark condensate; $\la \bar
ss\ra/\la\bar uu\ra\simeq 0.8$ measures the $SU(3)$ breaking of the quark condensate \cite{SNB}.  As shown
in the above formula,  the
$SU(3)$ breaking corrections are relatively small, and the four-quark condensate tends to decrease
the $\kappa$ mass \footnote{Using the instanton liquid model
\cite{SHURYAK}, the instanton contribution has been explicitly shown in \cite{DSN02} to be negligible.}.
Compared to the value of the
$\kappa$ mass given in Fig. \ref{fig: scalar1}b) using NWA, the finite width correction reduces the mass
by about 20 MeV, where we have used the width of 310 MeV from BES
\cite{BES}. One should mention that at the value of $\tau_0$, the OPE in powers of $\tau$ converges,
while the radiative corrections to the parton graph, though large in individual sum rules, remain small 
 in the ratio of moments ${\cal R}(\tau)$, as these corrections
tend to compensate each others, justifying the uses of the result at this relativley low scale.  From the previous
analysis, we deduce:
\beq
M_{a_0}\simeq 930~{\rm MeV}~~~~~~{\rm and}~~~~~~ M_{\kappa}\simeq 920~{\rm MeV}~,
\eeq
with about 10\% error, in good agreement with recent data \cite{PDG,BES,BEDIAGA}
\footnote{Previous sum rule analysis of the $\kappa$ parameters \cite{SNB} used as in put a value of $m_s$ and $t_c$ much higher than
here and gives higher value of the $\kappa$-mass.}.\\ 
{\bf The decay constants}\\
The decay constant $f_{a_0}$ of the $a_0$ normalized as:
\beq
\la 0\vert \partial_\mu V^\mu_{\bar ud}\vert a_0\ra\equiv \sqrt{2} f_{a_0}M^2_{a_0}~,
\eeq
in the same way as $f_\pi=92.4$ MeV has been estimated several times in the literature
\cite{SNB,FA0}:
\beq
f_{a_0}\simeq (1.6\pm 0.5)~{\rm MeV}~,
\eeq
where a better accuracy is claimed in \cite{MALT}.
Using $SU(3)$ symmetry and the almost degeneracy of the $a_0$ and $\kappa$ masses, we expect to have with
a good accuracy:
\beq
{f_{\kappa}\over f_{a_0}}\simeq {m_s-m_u\over m_d-m_u}\simeq 40~.
\eeq
{\bf The hadronic couplings}\\
The $a_0$ and $\kappa$ hadronic couplings have been obtained using either a vertex sum rule
\cite{PAVER,SNA0} or/and $SU(3)$ symmetry rotation \cite{BN}. The leading order vertex sum rule results
are:
\bea
g_{a_0K^+K^-}&\simeq& {8\pi^2\over 3\sqrt{2}}{m_s\la\bar ss\ra\over M^2_Kf_K}\ga 1-{2\over r}\dr\simeq
3~{\rm GeV}\nnb\\
{g_{a_0K^+K^-}\over g_{\kappa K^+\pi^-}}&\simeq& e^{-(M_K^2-m_\pi^2)\tau_0}{\ga 1-{2\over r}\dr}
\simeq 1.17~,
\eea
where we have used \cite{SNB}: $m_s\la\bar ss\ra\simeq -0.8 M^2_Kf^2_K$, $r\equiv \la \bar ss\ra/\la\bar
uu\ra\simeq 0.8.$ and $\tau_0\simeq 1$ GeV$^{-2}$. We expect an accuracy of about 20\% (typical for the
3-point fonction sum rules) for these estimates. Using the
$SU(3)$ relation: 
\beq
g_{a_0\eta\pi}\simeq \sqrt{2\over 3}g_{a_0K^+K^-}~
\eeq
one obtains:
\bea
\Gamma \ga a_0\rar\eta\pi\dr &\simeq& {\vert g_{a_0\eta\pi}\vert^2\over 16\pi M_{a_0}}\ga
1-{M^2_{\eta}\over M^2_{a_0}}\dr\nnb\\
&\simeq& 84~{\rm MeV}~, 
\eea
in agreement with the range of data from 50 to 100 MeV given by PDG \cite{PDG}.
Using the previous value of the $\kappa$ coupling, one can deduce:
\bea\label{eq: kappawidth}
\Gamma(\kappa\rar K\pi)\simeq {3\over 2}\Gamma(\kappa\rar K^+\pi-)\simeq 104~{\rm MeV}~,
\eea
which is about a factor 4 smaller than the present data \cite{PDG,BES}, but is a typical value for the width of a $\bar qq$ state.\\
{\bf The $\gamma\gamma$ widths}\\
The $\gamma\gamma$ width of the $a_0$ has been evaluated using vertex sum rules within the $\bar qq$ 
and four-quark assignements of this meson, with the result \cite{FA0,BN} :
\beq
\Gamma a_0(\bar qq)\rar \gamma\gamma\simeq (0.3\sim 2.1)~{\rm keV}~,
\eeq
and \cite{FA0}:
\beq
\Gamma a_0(4q)\rar \gamma\gamma\simeq (2\sim 5)\times 10^{-4}~{\rm keV}~,
\eeq
to be compared with the data of ($0.24\pm 0.08$) keV compiled in \cite{PDG}. Due to the inaccuracy of the QSSR predictions,
we shall definitely use, in the following, the measured value of the $a_0$ width as an input for our theoretical predictions of the $\bar
qq$ meson-width.\\
{\bf Concluding remarks}\\
$\b$ The previous analysis shows that the mass and widths of the $a_0(980)$ are well
described by a
$\bar qq$ assignement of this meson. \\
$\b$ A QSSR analysis of the four-quark
assignement \cite{JAFFE} of these states gives predictions which reproduce the experimental mass of the
$a_0(980)$ \cite{LATORRE,FA0,SNB}, like do the lattice calculations \cite{JAFFE2}. 
The result for the hadronic coupling
$g_{a_0K^+K^-}$ in the four-quark scenario depends crucially on the operators describing the $a_0$ and can range
from 1.6 GeV
\cite{MARINA} to (5--8) GeV \cite{FA0}. However, the prediction for the $\eta\pi$ can agree with
the data \cite{FA0} depending on the size of the operator mixing parameter. Therefore, an eventual
selection of the two approaches will be a precise measurement of $g_{a_0K^+K^-}$ or/and a lattice
measurement of the decay constant which should depend linearly on the light quark mass in the $\bar qq$
scheme, but is a constant in the four-quark one. \\
$\b$ Another problem arises when one computes the
$\gamma\gamma$ width using vertex sum rules. The ratio of the widths in the two approaches is
\cite{FA0} \footnote{We plan to come back to this point in a future work.}:
\beq
\Gamma_{4q\gamma\gamma}/\Gamma_{\bar qq\gamma\gamma}\approx (1\sim 2)\times 10^{-3}~,
\eeq
which is of the order of $(\alpha_s/pi)^2$, indicating that the four-quark assignement prediction is too small contrary to some claims in
the literature.
\\
$\b$ For the $\kappa$ meson, the $\bar qq$ assignement can reproduce quite well its {\it wrong} splitting with
the $a_0$, but fails to reproduce its large experimental width [about a factor 4 smaller (see Eq.
(\ref{eq: kappawidth})], while the four-quark one gives a width of about a factor 2 smaller \cite{MARINA}.
The failure of the two separate approaches ($\bar qq$ and four-quark assignement)  may suggest that the
 quark content of the $\kappa$(841) is more complex than na\"\i vely expected: it can be a mixing between a
$\bar qq$ and a four-quark states, or it can come from a large interference of the $\bar qq$ ground
state with non-resonant backgrounds. Further tests are needed for clarifying its nature.

\section{The $I=0$ bare scalar mesons}
\nin
The isoscalar scalar states are especially interesting in the framework of QCD since, in this anomalous $U(1)_V$
channel, their interpolating  operator is the trace of the energy-momentum tensor:
\beq
\theta_\mu^\mu=\frac{1}{4}\beta(\alpha_s) G^2+\sum_i [1+\gamma_m(\alpha_s)]
m_i\bar\psi_i\psi_i~,
\eeq
where $G^a_{\mu\nu}$ is the gluon field strengths, $\psi_i$ is the 
quark field; $\beta(\alpha_s)\equiv\beta_1\ga \alpha_s/\pi\dr+...$ and
$\gamma_m(\alpha_s)\equiv\gamma_1\ga \alpha_s/\pi\dr+...$ are respectively the QCD $\beta$-function and 
quark mass-anomalous dimension ($\beta_1=-1/2(11-2n/3)$ and $\gamma_1=2$ for $n$ flavours). In the chiral
limit
$m_i=0$,
$\theta_\mu^\mu$ is dominated by its gluon component $\theta_g$, like 
is the case of the $\eta'$ for the
$U(1)_A$ axial-anomaly, explaining why the $\eta'$-mass does not 
vanish like other Goldstone bosons for $m_i=0$. In
this sense, it is natural to expect that these $I=0$ scalar states 
are glueballs/gluonia or have at least a strong
glue component in their wave function. This gluonic part of $\theta^\mu_\mu$ should be identified with the $U(1)_V$ term \cite{DIVECCHIA} 
in the expression of the effective lagrangian based on a $U(3)_L\times U(3)_R$ linear realization of chiral symmetry (see e.g.
\cite{KYOTO,BLACK}). \\
{\bf Unmixed \boldmath$I=0$ scalar $\bar qq$ mesons}\\
We shall be concerned with the mesons $S_2$ and $S_3$ mesons associated respectively to the quark
currents:
\bea 
J_2=m:{1\over \sqrt{2}}(\bar uu+\bar dd):~~~{\rm and}~~~J_3=m_s:\bar ss:~. 
\eea
From the good realization of the $SU(2)$ flavour symmetry ($m_u=m_d$ and $\la\bar uu\ra=\la\bar dd\ra)$, 
one expects a degeneracy between the $a_0$ and $S_2$ states:
\beq
M_{S_2}\simeq M_{a_0}\simeq 930~{\rm MeV}~,
\eeq
while its hadronic coupling is \cite{BN,SNG}:
\bea\label{eq: s2pipi}
g_{S_2\pi^+\pi^-}\simeq {16\pi^3\over 3\sqrt{3}}\la\bar uu\ra \tau_0 e^{M^2_2\tau_0/2}\simeq 2.46~{\rm GeV}~.
\eea
corresponding to \footnote{We use
the normalization:
$$
\Gamma(\sigma_B\rar\pi\pi)={3\over 2}\frac{|g_{\sigma_B\pi^+\pi^-}|^2}
{16\pi M_{\sigma_B}}\ga{1-\frac{4m^2_\pi}{M^2_{\sigma_B}}}\dr^{1/2}.
$$.}:
\bea
&&\Gamma (S_2\rar \pi^+\pi^-)\simeq 120~{\rm MeV},
\eea
Using $SU(3)$ symmetry, one can also deduce:
\beq\label{eq: s2kk}
g_{S_2K^+K^-}\simeq {1\over 2}g_{S_2\pi^+\pi^-}\simeq 1.23~{\rm GeV}~.
\eeq
The $S_2$ $\gamma\gamma$ width can be deduced from the one of the $a_0(\bar qq)$ obtained previously, through the
non-relativistic relation (ratio of the square of quark charges):
\beq\label{eq:gamma}
\Gamma_{S_2\rar\gamma\gamma}\simeq {25 \over 9} \Gamma_{a_0\rar\gamma\gamma}\simeq (0.7\pm 0.2)~{\rm keV}~.
\eeq
The mass of the mesons
containing a strange quark is predicted to be \cite{SNG}:
\beq
M_{S_3}/M_{\kappa}\simeq 1.03\pm 0.02~~~~\Lrar M_{S_3}\simeq 948~{\rm MeV}~,
\eeq
if one uses $M_{\kappa}=920$ MeV \footnote{In \cite{SNG} a higher value has been obtained because one has
used as input the experimental mass $K^*_0=1430$ MeV.}, while its coupling to $K^+K^-$
is 
\cite{SNG}:
\bea\label{eq: s3kk}
&&g_{S_3K^+K^-}\simeq (2.7\pm 0.5)~{\rm GeV}~.
\eea
These results suggest that the na\"\i ve $\bar qq$ assignement of the $\sigma(600)\equiv S_2$ does not fit the data.
\\
{\bf Gluonia masses and decay constants}\\
$\b$ These states have been explicitly analyzed  in \cite{SNG0,VENEZIA,SNG} using QSSR of the two-point correlator:
\beq
\psi_s(q^2)=16i\int d^4x e^{iqx}\la 0|{\cal T}\theta^\mu_{\mu}(x)\theta^\mu_{\mu}(0)^\dagger |0\ra~,
\eeq
in the chiral limit $m_q=0$, for the observables:
\beq
{\cal L}_n(\tau)=-{\int_0^\infty dt~t^n e^{-t\tau} {\rm Im}\psi_s(t)}~, 
\eeq
and:
\beq
{\cal R}_{n,n+1}(\tau)=-{d\over d\tau}\log{\int_0^\infty dt~t^ne^{-t\tau} {\rm Im}\psi_s(t)}~,
\eeq
where $n=-1,~0,~1,~2$. For $n=-1$, the sum rule is sensitive to the subtraction constant:
\beq\label{eq: psi0}
\psi_s(0)=-16{\beta_1\over \pi}\la \alpha_s G^2\ra~,
 \eeq
 fixed from LET \cite{NSVZ}, where $\la \alpha_s G^2\ra=(0.07\pm 0.01)$ GeV$^4$ \cite{SNG2,SNB}.
One has found in \cite{VENEZIA} that, due to $\psi_s(0)$, the subtracted sum rule (SSR) ${\cal L}_{-1}$ is more
weighted by the high-energy behaviour of the spectral integral than the unsubtracted sum rule (USR) ${\cal
L}_{0,1,2}$, which motivated the introduction of 2 resonances (below and above 1 GeV) for solving the controversial
results obtained in the past. The results of the analysis using the standard OPE by retaining higher order PT series and
the lowest dimension condensates are \cite{SNG0,VENEZIA,SNG}:
\bea
f_{\sigma_B}&\simeq& (884\pm 116)~{\rm MeV}~, \nnb\\
 f_G&\simeq& (390\pm 145)~{\rm MeV}~,
\eea
corresponding to:
\beq\label{eq: massglue} 
M_{\sigma_B}\simeq 1~{\rm GeV}~~~{\rm and}~~~ M_G\simeq (1.5\pm 0.2)~{\rm GeV}.
\eeq
$M_{\sigma_B}$ has been obtained in \cite{SNG0} using a least square fit of ${\cal R}_{0,1}$. Its decay
constant has been obtained using a least square fit of ${\cal R}_{0,1}$ or/and a stability criterion of ${\cal
L}_0$ \cite{VENEZIA} \footnote{A similar analysis in the $U(1)_A$ channel has given an estimate of the $\eta'$
parameters and of the $U(1)_A$ topological charge \cite{SHORE,SNG0}.}. The mass of the 2nd
resonance has been fixed from
${\cal R}_{0,1}$
\cite{SNG} and its decay constant comes from ${\cal L}_{-1}$ \cite{VENEZIA}. We shall see that
$M_G$ is the one which can be compared with the present lattice value of about 1.6 GeV in the quenched
approximation \cite{PEARDON}, while the
$\sigma_B$ mass will be a $\eta'$-like meson expected to couple strongly with Goldstone boson pairs ({\it huge OZI violation})
\cite{VENEZIA,SNG} and playing a role in the saturation of the $U(1)_V$ two-point correlator subtraction
constant $\psi_s(0)$.  It can
only be tested using lattice with dynamical fermions (see e.g.
\cite{KYOTOG} for the inclusion of the disconnected part of the scalar propagator). \\
$\b$ One can also notice that a possible effect of the radial excitation of the $\sigma$ can be obtained by
matching the radial excitation contribution with the QCD continuum. Assuming its mass to be around 1.4 GeV, one
can deduce \cite{VENEZIA}:
\beq
f_{\sigma'_B}\leq (139\sim 224)~{\rm MeV}~,
\eeq
while a weaker bound of about 500 MeV has been allowed in \cite{SNG}.\\
$\b$ The effect of
the
$1/q^2$ term, to the previous results, which is an alternative of the direct instanton contribution has been shown
to be small \cite{CNZ}, though this term is necessary for solving the sum rule scale hierarchy of the gluonia
channels compared to the usual $\bar qq$ mesons. \\
$\b$ A recent QSSR analysis of the same
gluonium correlator using Gaussian sum rules and including instantons \cite{STEELE} confirms the previous mass values
obtained in Eq. (\ref{eq: massglue}), but not the results in ref.
\cite{FORKEL,STEELE2}, where it is argued that the presence of the direct instantons solve the controversial results noticed in
\cite{VENEZIA,SNG} between the subtracted $n=-1$ sum rule with the other $n\geq 0$ unsubtracted ones , without the need of two
resonances. In our normalization, the results in \cite{FORKEL} are:
\beq
M_S=(1.25\pm 0.2)~{\rm GeV},~~~f_S= (3\pm 0.3)~{\rm GeV}~.
\eeq
The mass value \footnote{It is an upper bound in \cite{STEELE2}.} does not contradict the ones in Eq. (\ref{eq: massglue}) as it
is about the mean value of the two resonances ones, while the decay constant leads to \cite{NSVZ,VENEZIA,SNG}:
\beq
{B}(J/\psi\rar S\gamma) \simeq 1.5\times 10^{-2}~,
\eeq
which is about 10 times higher than the one of the $J/\psi\rar f_2(1.24)\gamma$ and which is already excluded by the 
BES \cite{BES} and some other data. This fact may signal some eventual internal inconsistencies in the treatment of the instanton
contributions.
\\
$\b$ An upper bound on the gluonium mass has been also obtained in \cite{SNG}:
\beq
M_G\leq (2.16\pm 0.22)~{\rm GeV}~,
\eeq
using the positivity of the moment ${\cal R}_{1,2}$. This bound has been strengthened in \cite{FORKEL} to 1.7 GeV,
which cannot be an absolute bound because of the inclusion of the QCD continuum model and of its related uncertainties for its
derivation.
\\
{\bf Gluonia widths to \boldmath $\pi\pi$}\\
$\b$ 
For this purpose, we consider the gluonium-$\pi\pi$ vertex:
\beq
V(q^2)=\la\pi_1|\theta^\mu_\mu|\pi_2\ra,~~~~~q=p_1-p_2~,
\eeq
where:
$V(0)=2m^2_\pi~$. 
In the chiral limit $(m^2_\pi \simeq 0)$, 
the vertex obeys the dispersion relation:
\beq
V(q^2)=\int_0^\infty \frac{dt}{t-q^2-i\epsilon}
~\frac{1}{\pi}\mbox{Im} V(t)~,
\eeq
which gives the 1st NV sum rule \cite{VENEZIA}:
\beq
\frac{1}{4}\sum_{S\equiv\sigma_B,\sigma'_B,G}g_{S\pi\pi}\sqrt{2}f_S \simeq 0~.
\eeq
Using the fact that $ V^{\prime} (0)=1$ \cite{NSVZ2}, one obtains 
the 2nd NV sum rule \cite{VENEZIA}:
\beq
\frac{1}{4}\sum_{S\equiv\sigma_B,\sigma'_B,G}g_{S\pi\pi}\sqrt{2}f_S/M^2_S=1~.
\eeq
To a first approximation, we assume a $\sigma_B$-dominance in the 2nd NV sum rule, while in the 1st sum rule, there is
a matching between the $\sigma_B$ and all higher mass glueball contributions, which we replace by an effective $\sigma'$
mass of about 1.4 GeV. Then, one obtains:
\bea\label{eq: sigmapipi}
g_{\sigma_B\pi^+\pi^-}\simeq g_{\sigma_BK^+K^-} \simeq (3.2\sim 6.8)~{\rm GeV}~.
\eea
A complete matching in the 1st NV sum rule would lead to an effective coupling:
\bea
g_{\sigma'_B\pi^+\pi^-}\approx 12~{\rm GeV}~,
\eea
when using $f_\sigma'\simeq 500$ MeV. An interpretation of this value of $g_{\sigma'_B\pi\pi}$ is unclear but it is
expected to parametrize all higher states contributions to the 1st sum rule. If one uses a resonance mass of about 1.4
GeV, one  would obtain a very broad $\sigma'_B$ which can mimic the {\it red dragon} proposed earlier
\cite{OCHS}, but cannot manifest as a peak, making its identification with a true resonance difficult. Different strategies
for extracting the $f_0(1370)$ parameters from the continuum background has been discussed in \cite{ANISO} leading to
a width of the $f_0$ compiled in PDG \cite{PDG}. For definiteness, we shall use in the following, the value of
$\Gamma(\sigma'\rar\pi\pi$ about 250 MeV obtained in \cite{ANISO} within a factor two accuracy. This leads to:
\beq
g_{\sigma'_B\pi^+\pi^-}\simeq g_{\sigma'_BK^+K^-}\simeq (4.2\pm 0.7)~{\rm GeV}~.
\eeq
{\bf \boldmath $\sigma_B$ as an $\eta'$-like meson}\\ 
$\b$ One can check that the LET for $\psi_s(0)$ given in Eq. (\ref{eq: psi0}) is almost saturated by the contribution
of the lowest mass
$\sigma_B(1.)$ meson:
\beq
\psi_s(0)\simeq 2M^2_{\sigma_B} f_{\sigma_B}^2 \simeq 1.5~ {\rm GeV^4}~,
\eeq
compared to the LET value of 1.6 GeV$^4$. This property is very similar to the contribution of the $\eta'$ in the 
topological charge (subtraction constant of the anomalous $U(1)_A$ gluonium two-point correlator \cite{WITTEN}),which explains why
it is not degenerated with the pion at finite $N_c$.\\
$\b$ The $\sigma_B$ large coupling to pseudoscalar pairs, through OZI violating process, can also be compared with the affinity of the
$\eta'$ to couple to ordinary mesons, making it as an ambidextre gluonium-meson state \\
$\b$ Like in the $U(1)_A$ sector, the quenched lattice simulations obtain a higher glueball mass which is not the $\eta'$ mass. The
$\eta'$-mass and decay constant are only measured from a lattice calculation of the $U(1)_A$ topological charge \cite{GIACOMO}. We 
expect  that the same situation occurs in the $U(1)_V$ channel, where a quenched lattice gives a scalar gluonium mass
of about 1.5 GeV, while the $\sigma_B$ parameters can be obtained from the measurement of the scalar correlator subtraction constant
$\psi_s(0)$ including dynamical fermions.
\\
{\bf \boldmath $G(1.5)$ widths into \boldmath $\eta'\eta',~\eta\eta'$ and $\eta\eta$}\\
$\b$ \nin
Analogous low-energy theorem \cite{VENEZIA} gives:
\beq
\la \eta_1|\theta^\mu_\mu|\eta_1\ra = 2M^2_{\eta_1},
\eeq
where $\eta_1$ is the unmixed $U(1)$ singlet state of mass
$M_{\eta_1}\simeq $ 0.76 GeV \cite{WITTEN}.
Writing the dispersion relation for the vertex, one obtains the NV 
sum rule:
\beq
\frac{1}{4}\sum_{S\equiv\sigma_B,\sigma'_B,G}g_{S\eta_1\eta_1}\sqrt{2}f_S=
2M^2_{\eta_1},
\eeq
which, by assuming a $G$-dominance of the
vertex sum rule, leads to:
\beq \label{coup}
g_{G\eta_1\eta_1}\approx (1.2\sim 1.7)~\mbox{GeV}.
\eeq
Introducing the ``physical" $\eta'$ and $\eta$ through:
\bea 
\eta'\sim \cos\theta_P \eta_1-\sin\theta_P \eta_8\nnb\\
\eta\sim \sin\theta_P \eta_1+\cos\theta_P \eta_8,
\eea
where \cite{PDG,GILMAN}
$\theta_P\simeq -(18\pm 2)^\circ $ is the pseudoscalar mixing angle, one can deduce:
\beq
\Gamma(G\rar\eta'\eta)\simeq (5\sim 10)~{\rm MeV}.
\eeq
The previous scheme is also known to predict (see NV and \cite{GERS}):
\beq
r\equiv \frac{\Gamma_{G\eta\eta}}{\Gamma_{G\eta\eta'}}\simeq 0.22,~~~~~
g_{G\eta\eta}\simeq \sin\theta_Pg_{G\eta\eta'},
\eeq
compared with the GAMS data \cite{PDG} $r\simeq 0.34\pm 0.13$.
This result can then suggest that the $G(1.6)$ seen by the GAMS group is a 
pure gluonium, which
is not the case of the particle seen by Crystal Barrel \cite{PDG}
which corresponds to $r\approx 1$.\\
{\bf\boldmath Gluonia widths into \boldmath $4\pi$}\\
\nin
Within our scheme, we expect that the $4\pi$ are mainly $S$-waves initiated 
from the decay of pairs of $\sigma_B$. Using:
\beq
\la \sigma_B|\theta^\mu_\mu|\sigma_B\ra = 2M^2_{\sigma_B},
\eeq
and writing the dispersion relation for the vertex, one obtains the sum 
rule:
\beq
\frac{1}{4}\sum_{i=\sigma_B,\sigma'_B,G}g_{S\sigma_B\sigma_B}\sqrt{2}f_S=
2M^2_{\sigma_B}.
\eeq
We use $M_{\sigma_B}\simeq 0.6\sim 1$ GeV, $M_{\sigma'_B}\simeq 1.4$ GeV and the observed $f_0(1.37)$
width into $4\pi$ of about $(106\sim 250)$ {MeV} \cite{PDG} ($S$-wave part). Neglecting, to a first approximation, 
the $\sigma_B$ contribution to the sum rule, we can deduce:
\beq
|g_{G\sigma_B\sigma_B}|\approx 1.3\sim 3.7~\mbox{GeV}~,
\eeq
where the first (resp. second) value corresponds to $M_{\sigma_B}\simeq 0.6$ GeV (resp. 1 GeV). This leads to the width
into $\sigma(600)\sigma(600)$ of about  $(7-55)$ {MeV},  much larger than the one into $\eta\eta$ and $\eta\eta'$.
This feature is satisfied
by the $G(1.5)$ state seen by GAMS, Crystal Barrel and WA102 \cite{PDG}. However, the previous approaches show 
the consistency in
interpreting the $G(1.5)$ seen at GAMS as an ``almost" pure gluonium state
(ratio of the $\eta\eta'$ versus the $\eta\eta$ widths), 
while the
state seen by the Crystal Barrel and WA102, though having a gluon component 
in its wave function,
cannot be a pure gluonium because of its prominent
decays into $\eta\eta$ and $\pi^+\pi^-$. \\
{\bf Gluonia widths into $\gamma\gamma$}\\
These widths have been derived in \cite{VENEZIA} by identifying the $\gamma\gamma$-glue-glue box diagram with the scalar $\gamma\gamma$
Lagrangian where the quarks in the internal have been taken to be non-relativistic. In this way, one has obtained:
\beq
g_{\sigma\gamma\gamma}\simeq {\alpha\over 60}\sqrt{2}f_\sigma M^2_\sigma \ga{\pi\over -\beta_1}\dr \sum_{u,d,s} Q_i^2/m_i^4~,
\eeq
where $Q_i$ and $m_i$ are the quark charge and constituent masses. This leads to:
\bea
\Gamma (\sigma_B\rar\gamma\gamma)\simeq (0.03\sim 0.08)~{\rm keV}~,\nnb\\
\Gamma (\sigma'_B\rar\gamma\gamma)\simeq (0.01\sim 0.03)~{\rm keV}~,\nnb\\
\Gamma (G\rar\gamma\gamma)\simeq (0.3\sim 0.6)~{\rm keV}~.
\eea
Alternatively, one can use the trace anomaly to order $k^4$ in order to deduce:
\beq
\la 0|{1\over 4}\beta(\alpha_s) G^2|\gamma_1\gamma_2\ra\simeq -\la 0|\ga {\alpha\over 3\pi}\dr 
R F_1^{\mu\nu}F_2^{\mu\nu}|\gamma_1\gamma_2\ra~,
\eeq
where $R\equiv 3\sum _i Q^2_i$, $\alpha$ is the QED coupling. This relation gives:
\beq
{\sqrt{2}\over 4}\sum_{\sigma,...}f_ig_{i\gamma\gamma}\simeq \ga{\alpha\over 3\pi}\dr R~.
\eeq
From this relation and using the previous values of $f_\sigma$ and $f_\sigma'$, one can deduce:
\beq
\Gamma_{G\rar\gamma\gamma}\simeq (1\sim 6)~{\rm keV}~,
\eeq
which is quite inaccurate but still consistent with the previous determination.\\ 
{\bf Comments}\\
Comparing the above results, especially the predicted widths, with the experimentally
observed candidates, it is likely that the
$\sigma$ and some of its radial excitations have a lot of glue in their wave functions. As a consequence, a
quarkonium-gluonium decay mixing scheme~\footnote{This has to be contrasted with the small mass-mixing coming
from the off-diagonal two-point function \cite{PAK}.} has 
been proposed in the $I=0$ scalar sector.
\cite{BN,SNG}, for explaining the observed spectrum and widths of the 
possibly wide $\sigma (< 1$ GeV) and the narrow $f_0(980)$ \footnote {We shall not consider in our analysis the recent result of
\cite{MIKE} from
$\gamma\gamma\rar
\pi^0\pi^0$ where the resulting $\sigma\rar\gamma\gamma$ width of about 4 keV is much bigger than generally expected and which needs to
be confirmed by some other data. We plan to come back to this point in the future.}.

\section{\boldmath Meson-gluonium mixing below 1 GeV}
\nin
$\b$ BES data suggest that the $\sigma$(600) is produced in the OZI forbidden $J/\psi\rar\phi \pi^+\pi^-$
process \cite{BES}, which can indicate the large amount of glue in its wave function. Its production from the
OZI allowed $J/\psi\rar\omega \pi^+\pi^-,~K^+K^-$ processes, is expected to be due to its quark component, while the relative
small branching ratio in the OZI allowed $J/\psi\rar\phi K^+K^-$ process  relative to  $J/\psi\rar\phi  \pi^+\pi^-$ can
be due to an interference between the $K^+K^-$ amplitude from the gluon and quark components of the $\sigma$.\\ 
$\b$ In the same way, the $f_0(980)$ is produced in the OZI violating $J/\psi\rar\phi \pi^+\pi^-,$ \cite{BES} and
$D_s\rar\pi^-\pi^+\pi^+$ \cite{BEDIAGA} processes which may also indicate its gluonium component, while its production
from  $J/\psi\rar\phi K^+K^-,$ can signal a strong $\bar ss$ component in its wave function. We shall keep in mind these
results for building the mixing scheme.\\
{\bf The meson-gluonium mixing scheme}\\
$\b$ We assume that the observed states come from the mixing between the gluonium $\sigma_B$ and quark $S_2\equiv
{1/\sqrt{2}}(\bar uu+\bar dd)$ and
$S_3\equiv \bar ss$ bare states \footnote{In our approach, we first calculate the real part of the masses of
these hypothetical states and deduce their widths using vertex sum rules. These states would correspond to the bare
states in the
$K$-matrix formalism (see e.g. \cite{ANISO}). Due to the large error in our mass
predictions, we neglect, to a first approximation, some possible shifts on the masses which can be induced by the decay
processes mentioned in \cite{ANISO}.}:
\bea\label{eq: mixing1}
\ga \bear{c}
\sigma\\
f_0\\
\eear\dr =
\ga\bear{cc}
\cos{\theta_S}&\sin{\theta_S}\\
-\sin{\theta_S}&\cos {\theta_S}\\
\eear\dr\ga \bear{c}
\sigma_B\\
S_2+\phi_S S_3
\eear\dr\nnb
\eea
where $\phi_S=1/\sqrt{2}$ for a $SU(3)$ singlet and $-\sqrt{2}$ for an SU(3) octet.\\
$\b$ In \cite{BN}, the mixing angle $\theta_S$ has been fixed from the analysis of the predicted decays of the
hypothetical bare states $S_2,~\sigma_B$ and of the observed meson $f_0(980)\rar\gamma\gamma$. Using the
predictions:
$\Gamma(S_2\rar\gamma\gamma)$ in Eq. (\ref{eq:gamma}), 
$\Gamma(\sigma_B\rar\gamma\gamma)\simeq 0.03$ keV \cite{VENEZIA}, and the recent data
$\Gamma(f_0(980)\rar\gamma\gamma)\simeq (0.4\pm 0.1)$ keV
\cite{PDG}, one can deduce:
\beq
{\theta_S}\simeq (45\pm 15)^0~,
\eeq
where the $S_3\rar\gamma\gamma$ width is suppressed as (2/25) compared to the one of the $S_2$.\\
$\b$ Using, for definiteness, as inputs the theoretical predictions given in Eqs (\ref{eq: s2pipi}) and (\ref{eq:
sigmapipi}),
we predict the couplings:
\bea
g_{\sigma\pi^+\pi^-}&\simeq&(5.3\pm 1.8)~{\rm GeV}~, \nnb\\
g_{f_0\pi^+\pi^-}&\simeq&(1.8\pm 1.3)~{\rm GeV},
\eea
in reasonnable agreement with the ones from the data:
\bea
g^{exp}_{\sigma\pi^+\pi^-}&\simeq&(3.0\pm 1.5)~{\rm GeV}~, \nnb\\
g^{exp}_{f_0\pi^+\pi^-}&\simeq&(1.5\pm 0.3)~{\rm GeV}~,
\eea
corresponding to:
\bea
\Gamma^{exp}(\sigma(600)\rar \pi\pi)&\approx& 481~{\rm MeV}~,\nnb\\
\Gamma^{exp}(f_0(980)\rar \pi\pi)&\approx& (70\pm 30) ~{\rm MeV}~,
\eea 
$\b$ In order to predict the mixing parameter $\phi_S$, we fit the experimental $f_0K^+K^-$ width from BES
\cite{BES}:
\beq
{g_{f_0K^+K^-}/ g_{f_0\pi^+\pi^-}}\simeq 2.05\pm 0.15~,
\eeq
and we use the theoretical predictions given in Eqs (\ref{eq: s2pipi}), (\ref{eq: s3kk}) and (\ref{eq: sigmapipi}).
Then, we obtain:
\beq
\phi_S\simeq 3.0~,~~~~~{\rm and}~~~~~{g_{\sigma K^+K^-}/ g_{\sigma\pi^+\pi^-}}\simeq 2~.
\eeq
Further data are needed for improving and testing this result.\\
$\b$ One should note that using only the constraint from the $f_0\rar \pi^+\pi^-$ and $f_0\rar K^+K^-$ hadronic widths,
one would obtain:
\beq\label{eq: scheme2}
\theta_S=16^0~,~~~~{\rm and}~~~~~~~\phi_S\simeq -1.4~,
\eeq
indicating that the $\sigma$ is an almost pure gluonium and the $f_0$ a $\bar qq$ $SU(3)$ octet, which is similar
to the scheme in \cite{OCHS}. However, the result in Eq. (\ref{eq: scheme2}) would give a too high value of $f_0\rar\gamma\gamma$ and
does not explain the OZI violating production of the $f_0$ in $J/\psi$ and $\phi$ radiative decays. \\
{\bf Comments on alternative approaches }\\
$\b$  One should note that a four-quark QSSR analysis gives \cite{MARINA}:
\bea\label{eq: 4quark}
g_{f_0K^+K^-}&\simeq& g_{a_0K^+K^-}\simeq (1.6\pm 0.1)~{\rm GeV}\nnb\\
 g_{f_0\pi^+\pi^-}&\simeq&(0.47\pm 0.05)~{\rm GeV}~, 
\eea
where the absolute values differ from the ones given in \cite{ACHASOV2} and then question the realibility of the
results obtained there.
Therefore, lattice calculations of these couplings become mandatory. Eq. (\ref{eq: 4quark})  leads to:
\bea
\Gamma(f_0(980)\rar \pi\pi)&\approx& 7 ~{\rm MeV}~,
\eea 
which is too small compared with the range $(40\sim 100)$ MeV given by the data \cite{PDG}.\\
$\b$ Alternative
approaches based on $\bar KK$ loop, $\bar KK$ molecules and four-quark states can predict value of
$\Gamma(f_0(980)\rar\gamma\gamma)$ in agreement with the data \cite{ESCRIBANO},\cite{POLOSA}, but, most of them, do not give a
satisfactory prediction for the $f_0\rar\pi^+\pi^-$ width and their production from OZI violating decays. On the other, it would be
interesting to see the connection of these effective approaches with the quark-gluon picture used here. \\
{\bf Tests from {\boldmath$J/\psi$ and $\phi$} radiative decays.}\\ 
$\b$ These decays are known to be a gluonium filter.
The production of a gluonium $S$ from $J/\psi$ radiative decays can be approximated by \cite{NSVZ2}:
\bea
&\Gamma(J/\psi\rar\gamma S)\simeq \frac{\alpha^3\pi}{\beta_1^2 656100}
\ga\frac{M_{J/\psi}}{M_c}\dr^4\ga\frac{M_{S}}{M_c}\dr^4\nnb\\ 
&\frac{\ga 1-M^2_S/M^2_{J/\psi}\dr^3}{\Gamma(J/\psi\rar e^+e^-)}f^2_S~,
\eea
where $M_c\simeq 1.5$ GeV is the charm constituent quark mass. In our scheme, the $\sigma$ is mostly a gluonium.
Therefore, one expects the branching ratio:
\beq
B(J/\psi\rar\gamma \sigma)\approx 19\times 10^{-5}~.
\eeq
Extrapolating the previous exppression to the $\phi$-meson and using $M_s\simeq 500$ MeV, one obtains:
\beq
B(\phi\rar\gamma \sigma)\approx 12\times 10^{-5}~,
\eeq
which, despite the crude approximation used, compares quite well with the KLOE \cite{KLOE} data:
\beq
B(\phi\rar\gamma \pi^0\pi^0)\approx \ga 10.9\pm 0.3\pm 0.5\dr\times 10^{-5}~.
\eeq
{\bf Tests from {\boldmath$D_{(s)}$} semileptonic decays}\\
\nin
This section has been discussed in details in \cite{DSN02} and will be only sketched in
the following.\\
$\b$ {\boldmath$S_2(\bar uu+\bar dd)$ \bf meson productions}\\
\nin
If the scalar mesons were simple $\bar qq$ states, the 
semileptonic decay width
could be calculated quite reliably using QSSR, where the relevant 
diagram is a quark loop triangle.
Several groups \cite{Dosch:2002rh} predict 
all form factor to be:
$f_+(0)\approx 0.5~,$
yielding, for $M_{S_2}\simeq 600$ MeV, a decay rate:
\beq
\Gamma(D\rar S_2 l\nu)=(8\pm 3)10^{-16} ~{\rm GeV}~,
\eeq
which is, unfortunately, even in high stastistics
experiments, at the 
edge of observation since the
decays into an isoscalar are CKM-suppressed due to the $c$-$u$ transition at
the weak vertex.\\
$\b$ {\bf Scalar gluonium or/and \boldmath $\bar ss$ productions}\\
\nin
Semi-qualitative but model independent results for the production of gluonium
have been given in \cite{DSN02} (see also \cite{KISSLINGER}):
\\
{\bf --} The only way to obtain a non-CKM suppressed isoscalar is to look at the
semileptonic decay of the $D_s$-meson, where the light
quark is a strange one and an isoscalar $s \bar s$ or/and gluonium state can be
formed. \\
{\bf --} If the $\bar ss$ state is relatively light ($<1$ GeV), which might be 
the natural partner of the ($\bar
uu+\bar dd$) often interpreted  to be a $\sigma (600)$ in the 
literature, then, one should produce a
$ K\bar K$ pair through the isoscalar $\bar ss$ state. The
non-observation of this process will disfavour the
$\bar qq$ interpretation of the $\sigma$ and $f_0$ mesons. \\
{\bf --} If a gluonium state is formed it will decay with even strength 
into $\pi\pi$ and a $K\bar K$ pairs. Therefore a gluonium
formation in semileptonic $D_s$ decays should result in the decay patterns:
\beq
D_s\rar \sigma_B\ell \nu\rar \pi \pi \ell \nu ~~~~~ 
D_s\rar\sigma_B\ell \nu\to K \bar K  \ell \nu~,
\eeq
with about the same rate up to phase space factors.
The observation of the semileptonic $\pi\pi$ decay of the $D_s$ 
by E791 \cite{BEDIAGA} is a
sign for glueball formation \footnote{An alternative explanation assuming $f_0(\bar ss)$ and using $\bar KK$ loop
has been given in \cite{POLOSA}. However, the same assumption for $f_0$ but using QSSR leads to a negative
conclusion \cite{BEDIAGA2}.}.\\
{\bf --} Using,
e.g., the result in ~\cite{Dosch:2002rh}, the one for light $S(\bar qq)$ 
quarkonium production behaves as:
\beq
\Gamma[D_s\rar S_q~l\nu]\sim |V_{cq}|^2G^2_F M^5_c |f_+(0)|^2~.
\eeq
{\bf --} The $\sigma_B(gg)$ production, can be obtained from the $1/M_c$ 
behaviour of the $WW gg$ box diagram. Using dispersion techniques similar to 
the one used for
$J/\psi\rar \sigma_B \gamma$ processses \cite{NSVZ,VENEZIA,SNB}, one 
obtains, assuming a $D_s$ and
$\sigma_B$-dominances \cite{DSN02}:
\beq
\Gamma[D_s\rar\sigma_B~l\nu]\sim |V_{cs}|^2G^2_F  \frac{|\la 
0|\phi_S G^2|\sigma_B\ra|^2}{M_cM_\sigma^4},
\eeq
where $\la 0|\phi_S G^2|\sigma_B\ra $ is by definition 
proportional to $f_\sigma M^2_\sigma$. Then, one
deduces:
\beq
{\Gamma[D_s\rar\sigma_B(gg)~l\nu]\over \Gamma[D_s\rar S_q(\bar qq)~l\nu]}
\sim {1\over |f_+(0)|^2}\ga{f_\sigma \over M_c}\dr^2,
\eeq
which is $ {\cal O}(1)$ for $f_\sigma\simeq 0.8$ GeV.\\
$\b$ {\bf Comments}\\
These semi-quantitative results indicate that { 
the gluonium production rate can be of the same order as the
$\bar qq$ one contrary to the na\"\i ve perturbative expectation 
($\alpha_s^2$ suppression), which is a
consequence of the OZI-rule violation of the $\sigma_B$ decay.}~\footnote{Productions of the scalar mesons in $B$-decays have been
discussed in \cite{OCHS2}.} However, it also shows that, due to the (almost) universal 
coupling of the $\sigma_B$ to Goldstone boson pairs, one also
expects a production of the $K\bar K$ pairs, which can compete with 
the one from $\bar ss$ quarkonium state, and again renders more
difficult the identification of the such $\bar ss$ state if allowed 
by phase space. 
\section{Properties of the mesons above 1 GeV}
\nin
QSSR does not have a precise systematic framework for extracting the properties of the radial excitations, except
the approximate value of the mass indicated by the value of the QCD continuum threshold at which the mass of
the ground state has been optimized. In order to check this result, one often uses (if available) empirical
observations of the splittings between the radial excitations and the ground state, or/and arguments based on
the (linear) Regge trajectories. For this reason, the discussions which we shall give below will be very
qualitative. \\
{\bf The $I=1$ and $1/2$ mesons}\\
The $a_0(1450)$ and the $K^*_0(1430)$ are almost degenerated indicating the restoration of the $SU(3)$
flavour symmetry where the $SU(3)$ breakings behaves like $m_s^2/M_R^2$ and $m_s\la\bar ss\ra/M_R^4$ for the
radial excitations. These scalar radial
excitations are also almost degenerated with the pseudoscalar
$\pi(1300)$ and $K(1460)$ also indicating the restoration of the spontaneous breaking of the $SU(3)_L\times SU(3)_R$
chiral symmetry broken by the quark condensate at higher scale \footnote{Similar though qualitative arguments have been
given in \cite{GLOZMAN}.}. Within such observations, one can expect that these scalar states are $\bar qq$ states\footnote{Similar
conclusions have been also independently reached in \cite{KATAEV} using FESR.}. \\
{\bf The {\boldmath$I=0$ scalar mesons}}\\
There are proliferations of these states from the data \cite{PDG,BES}: $f_0(1370),~f_0(1500),~f_0(1710)$ and
$f_0(1790)$. Using the previous symmetry restorations, one may expect that the 1st radial excitation of the $(\bar
uu+\bar dd)$ and $\bar ss$ unmixed states should be in the range of 1400 MeV, which are in the vicinity of the 
$f_0(1370)$ and $f_0(1500)$ being good experimental candidates. The 2nd radial excitations are expected to be in the range
of the $\pi(1800)$ where good experimental candidates are the $f_0(1710)$ and $f_0(1790)$. Within these observations in
mind, we shall interpret the different data given by BES \cite{BES}:\\
$\b$ The {$f_0(1370)$ and $f_0(1710)$} are respectively produced
through the OZI forbidden 
$J/\psi\rar\phi \pi^+\pi^-$ and allowed $J/\psi\rar\omega K^+K^-,$ processes \cite{BES}, while the latter is also produced from the
glueball filter process  $J/\psi\rar\gamma K^+K^-$. These features can indicate that they can have an important gluonium component in
their wave function. To a first approximation and iteration, we assume that these two states result from
the mixing \footnote{Some other alternative mixing schemes above 1 GeV have been also proposed 
\cite{ANISO,SNG}, \cite{FAESSLER,OCHS}, \cite{CLOSE,VIJANDE}.} :
\bea\label{eq: mixing2}
\ga \bear{c}
f_0(1370)\\
f_0(1710)\\
\eear\dr =
\ga\bear{cc}
\cos{\theta'_S}&\sin{\theta'_S}\\
-\sin{\theta'_S}&\cos {\theta'_S}\\
\eear\dr\ga \bear{c}
\sigma'_B\\
S'_2+\phi'_S S'_3
\eear\dr\nnb
\eea
which is a replica of the mixing among ground states in Eq. (\ref{eq: mixing1}).
We fix the coupling of the $\bar qq$ radial excitations by assuming that they behave like the one of the pion and
$\pi(1300)$,  namely:
\beq
g_{S'_{2,3}\bar PP}\approx \ga{M_{S'_{2,3}}\over M_{S_{2,3}}}\dr^n g_{S_{2,3}\bar PP}~,
\eeq
where $P\equiv \pi,K$;  $n\simeq 2$. Then, we deduce:
\bea
g_{S'_{2}\pi^+\pi^-}&\simeq& 2g_{S'_{2}K^+K^-}\simeq 4.8~{\rm GeV}~,\nnb\\
g_{S'_{3}K^+K^-}&\simeq& 6.4~{\rm GeV}~.
\eea
Using the experimental input from BES \cite{BES}:
\bea
\Gamma(f_0(1370\rar \pi\pi)&\simeq& (265\pm 40)~{\rm MeV},\nnb\\
{\Gamma(f_0(1370\rar \bar KK)\over \Gamma(f_0(1370\rar \pi\pi)}&\simeq&(0.08\pm 0.08) ,
\eea
one can fix the two angles to be:
\bea
\theta'_S\simeq 75^0~~~~~~\phi'_S\simeq -0.55~.
\eea
The solution indicates that the $f_0(1370)$ contains more $(\bar uu+\bar dd)$ than a gluonium, while the $f_0(1710)$ 
contains more glue and $\bar ss$ than $(\bar uu+\bar dd)$, which is in line with general trends. From this analysis, we
predict:
\bea
g^{th}_{f_0(1710)\pi^+\pi^-}\approx 2.8~{\rm GeV}~,\nnb\\
g^{th}_{f_0(1710)K^+K^-}\approx 4.3~{\rm GeV}~,
\eea
which lead to the widths:
\bea
\Gamma[{f^{th}_0(1710)\rar\pi^+\pi^-}]\approx 90~{\rm MeV}~,\nnb\\
\Gamma[{f^{th}_0(1710)\rar K^+K^-}]\approx 176~{\rm MeV}~.
\eea
The results are of the order of the BES data,  which are respectively $\leq 16$ MeV and 
125 MeV \cite{BES}, but are not yet quite satisfactory. As the $f^{th}_0(1710)$ contains more glue after the mixing, it
is natural that it will mix with the glueball $G(1500)$ with the theoretical parameters obtained previously.\\
{$\b$} Therefore, in the 2nd step analysis, we consider that the observed $f_0(1710)$ and $f_0(1500)$ result from
the mixing between $f^{th}_0(1710)$ and $G(1500)$.\\
$\b$ {The} $f_0(1500)$
is produced in the gluonium filter process $J/\psi\rar\gamma \pi^+\pi^-$, and it is needed for improving
the fit of the OZI violating $J/\psi\rar\phi \pi^+\pi^-$ process, but has a small effect in the  $J/\psi\rar\phi
\bar KK$ process \cite{BES}. On the other, one expects from previous sections that the $G(1500)$ couplings to $\pi\pi$ and
$\bar KK$ are small as the vertex sum rule is almost saturated by the $\sigma_B$, but we do not have a precise
quantitative control of such couplings. However, one can check that the model cannot explain simultaneously the small
$f_0(1500)\rar K^+K^-$ and $f_0(1710)\rar\pi^+\pi^-$ widths. \\
$\b$ A solution to this problem may be given by the mixing of the previous two states with the $f_0(1790)$ and some 2nd
radial excitations of the $\bar qq$ states expected to be in this range of energy \footnote{Alternative explanation has been in
\cite{CHANOWITZ}, where it is argued that the glueball $f_0(1710)$ has chiral coupling to pairs of Goldstone bosons using
perturbative QCD arguments. However, the validity of this result has been questioned in \cite{CHINOIS} due to instanton
contributions.}. The
$f_0(1790)$ may contain more gluon  in its wave function as it is produced in the glueball filter channel
$J/\psi\rar\gamma
\pi^+\pi^-$ and in the OZI violating
$J/\psi\rar\phi \pi^+\pi^-$ process, while it also decays into $4\pi$. The smallness of the $f_0(1790)\rar K^+K^-$ width
may come from a destructive interference between the gluonium and $\bar qq$ states~\footnote{An alternative explanation of the
$f_0(1790)$ width is based on the restoration of chiral symmetry which favours the previous OZI violation process production of the
$f_0(1790)$ \cite{GLOZMAN}. However, the author does not explain the absence of its $K^+K^-$ decay.}.\\
\section{Conclusions}
\nin
Motivated by the various recent experimental progresses for producing scalar mesons, we have proposed new results 
and
updated previous predictions of \cite{SNG,VENEZIA,BN}:
\\
$\b$ The unusual {\it wrong} splitting between the $a_0(980)$ and $\kappa(900)$
being respectively a $\bar ud$ and $\bar us$ mesons can be understood from QSSR using the present value of
$\mb_s$.
\\
$\b$ The maximal meson-gluonium mixing
for the $\sigma(600)$ and $f_0(980)$ appears to be supported by the present data.\\
$\b$ The $a_0(\bar ud)$,
$\kappa(\bar us)$, $\sigma(600)$ and $f_0(980)$ mesons appear to complete the scalar nonet below 1 GeV.\\
$\b$ The productions of the $I$=0 mesons above 1 GeV through the OZI
violating
$J/\psi\rar
\phi\pi^+\pi^-
$,
$D_s\rar 3\pi$ decays and $J/\psi\rar \gamma S$ glueball filter processes may indicate that most of these $I=0$ mesons
have important gluonium in their wave functions, where a simple meson-gluonium mixing scheme can explain the general
features of the data. \\
$\b$ Our results suggest that the $f_0(1370)$ is mostly $\bar qq$, while the $f_0(1500),~f_0(1710)$ and
$f_0(1790)$ have significant gluonium component in their wave functions. Tests of these results can be provided by the
measurements of the pure gluonium $\eta'\eta$ and $4\pi$ specific $U(1)_A$ decay channels.\\ We conclude that, after about a 1/4
century study, we still remain with more questions  than answers on the true nature of scalar mesons.
\section{Acknowledgements}
\nin
It is a pleasure to thank Jean-Marc Richard for some comments on the preliminary draft.

\end{document}